# Comment on "Synthesis of rhenium nitride crystal with MoS$_2$ structure" [APL 100, 251910(2012)]

G. Soto

*Universidad Nacional Autónoma de México, Centro de Nanociencias y Nanotecnología*
*Km 107 carretera Tijuana-Ensenada, Ensenada Baja California, México.*

Kawamura *et. al.* [1] recently published an article about the synthesis of rhenium nitride with MoS$_2$-type structure. In summary, in this article the high-pressure high-temperature metathesis reaction between lithium nitride and rhenium pentachloride was applied to produce materials with platelet shapes. They report powder x-ray diffraction (XRD) patterns using Bragg-Brentano and Gandolfi geometries. All diffraction peaks where indexed to a hexagonal *P6$_3$/mmc* (194) space group, isostructural with MoS$_2$ and similar to ReB$_2$. The Rietveld structure analysis gave lattice constant of $a$ = 2.806 Å and $c$ = 10.747 Å. XPS analysis where used to confirm the Re-N bonding. The authors conclude that their synthesis "reversed the theoretical predictions". This conclusion was taken by the fact that the experimental structure parameters differ with some theoretical calculations of rhenium-nitrogen compounds in 1:2 stoichiometric ratio. While the reason for the discrepancy were not clarified, it is assumed that this type of synthesis often develop unusual metastable phases.

First, let me clarify that I consider the work of Kawamura as very valuable, especially from the experimental point of view, since they synthesized a new material. But I disagree with the assertion that these findings are in contrast to theoretical predictions. Recently I published an article where the positive-enthalpy ReN$_2$ compound is proposed, along with other diazenide compounds[2]. The structure is equivalent to the one proposed by Kawamura *et. al.*, however the structural parameters are very different. I think my results are quite correct. Considerable effort has been put the field of *ab initio* crystal structure predictions. Specially, the density functional theory (DFT) is its maturity age and gives very reliable structures. When there are discrepancies the most likely cause is a human mistake, which may be on the theoretical side, but also in the experimental part. In this case I think there is a poor interpretation of data on the experimental side. I will show using DFT, that Kawamura *et. al.* underestimated the concentration of nitrogen in his material. Their synthesis leads to something like rhenium azide, ReN$_3$, instead of rhenium diazenide, ReN$_2$.

As a first test, DFT shows that the structure as proposed by Kawamura is not in equilibrium, that is, can not be a metastable phase. The plot of total energy versus volume is show in figure 1. The slope *dE/dv* at the Kawamura's point is not zero. This means that there are no internal cancelations of forces; this can not be a metastable point. The energy of ReN$_2$ as was proposed by Kawamura *et. al.* is very high, see Table I. Unless there is an external stress, this atomic arrangement by itself should attain the structural parameters and atomic positions calculated by DFT in ref. 1, also given in Table I. Evidently, the diffraction pattern given by DFT at this composition will not match the experiment of Kawamura.

As a second test, DFT will show that the Kawamura structure is achievable, but with another rhenium to nitrogen stoichiometric ratio. Between the atoms of rhenium there is plenty of space to insert $N_3^-$ azide anions. Unfortunately I did not get access to the original XRD data, so I must have to generate the patterns for a Mo-K$\alpha$ radiation with the data offered as of Kawamura's structure description. Figure 2 (a) shows the generated pattern for ReN$_2$ with the proposed cell parameters; that correspond to the first structure of Table I. Next to it, figure 2 (*b*), is the generated pattern of ReN$_3$ constrained to the *P6$_3$/mmc* space group, it correspond to the second figure of Table I. DFT optimization of volume, *c*/*a* ratio, and internal free atom coordinates were performed recursively. I find that this structure of ReN$_3$ is close to those reported by Kawamura. However the central nitrogen in the *P6$_3$/mmc* space group is positioned in a strained state. When fully relaxed, the hexagonal symmetry is lost and the ensuing structure fit in the *Ama2* (40) space group. In Figure 2 (*c*) is the generated pattern of ReN$_3$ without being forced to the *P6$_3$/mmc* space group. ReN$_3$ in the *Ama2* space group is illustrated in the third row of Table I.

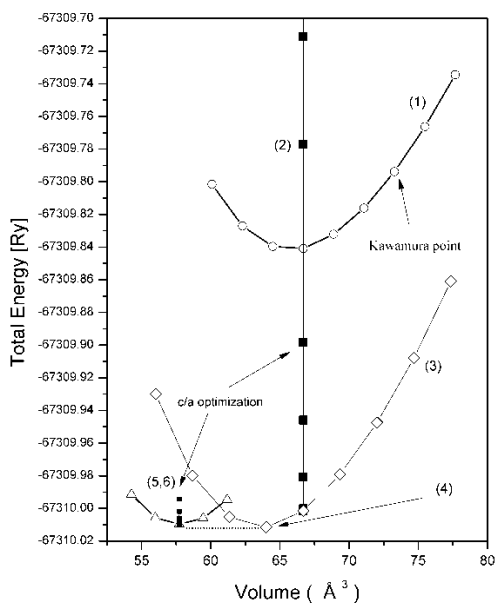

Figure 1.- Trajectory of structure optimization using the Kawamura's parameters as starting point.

The optimization of a structure like this is difficult because there are many free parameters and little symmetry. As we can see from this Figure 2, the diffraction patterns are very similar among them, so a numerical regression as is the Rietveld method cannot decide categorically. This is because Re is a strong x-ray scatterer, while N is moderate scatterer. Although the diffraction patterns generated show a probable cause of confusion, is not definitive proof that ReN$_3$ was synthesized in the Kawamura's experiment. The energies calculated by DFT are important to find out what structures are possible and what not. In Table I, $E_t$ is the total energy as it is calculated by the Wien2k code. It has contributions from the core, valence electrons and bonding energy. $E_A$ is the total energy minus the energy of forming atoms (rhenium in basal state and nitrogen in N$_2$-molecule); $E_A$ is the enthalpy at 0 K. $E_B$ is the formation energy starting from nitrogen in atomic state. For appraisal, the energies are given by atoms$^{-1}$. $E_A$ positive indicates the tendency to spontaneous dissociation to Re + $N_2$, while negative $E_B$ shows that these compound can be formed when atomic nitrogen is available in chemical reactions. $E_B$ negative indicate higher probability of being formed. The comparative analysis is given in the Table I. DFT shows that $ReN_3$ is likely, while $ReN_2$ with the Kawamura parameters is not. The diffraction pattern of $ReN_3$ is close to his experiment. While the energy comparison shows higher probability for $ReN_3$, the conclusive evidence that the Kawamura synthesis gave rise to azide-like compound should be in the XPS measurements. I want to see the region around the N-2$s$ transition. If I am correct, there should be a splitting ($\Delta E \sim 5$ eV) between molecular and atomic states, as the density of states calculated by DFT allows seeing in Figure 3. The lowest bands in $ReN_3$ derive $N_3^-$ molecular orbitals[3], states provided mainly by the central atom of the azide ion, while the upper portion corresponds mainly to states from the N atoms shaping bonds to Re. Now Kawamura has the floor. Let us to see your XPS measurements.

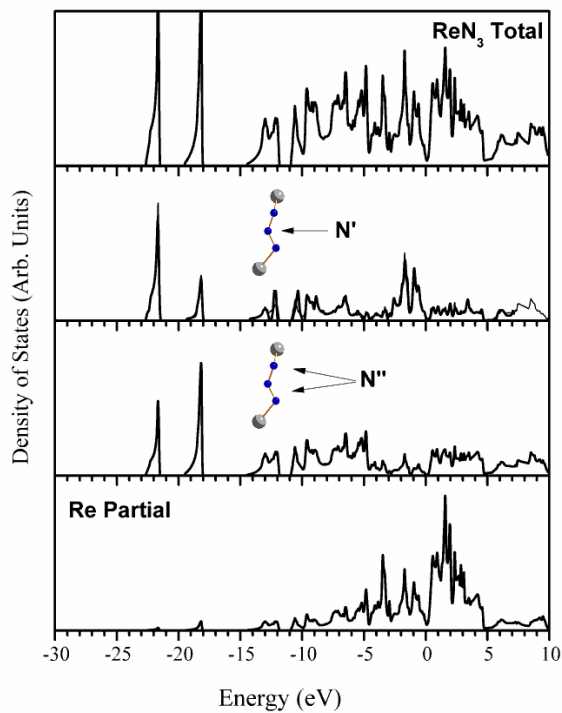

Figure 3. - Calculated density of states for rhenium azide. In the top trace is the total DOS of ReN3. The bottom trace is the contribution of the Re atoms. In the central plots are the projections from the internal and external nitrogen atoms from the azide group.

In conclusion, the compound synthesized by Kawamura is $ReN_3$, or a compound of similar composition. What Kawamura did is to find the Re atomic positions of a compound where the nitrogen concentration remains unknown. While these are positive-enthalpy compounds, like silver azide ($AgN_3$), mercury azide ($Hg_2N_6$), and barium azide ($Ba(N_3)_2$), their synthesis is feasible starting from atomic nitrogen. The theoretical calculations are, without doubt, a powerful tools that can be used by experimentalist (as is my case) to give better interpretations to their results. Despite the DFT progress there is still much work to do, especially in trustworthy protocols to predict the structures of unknown materials.

As a final comment, Kawamura *et. al.* says in its paper that "these nitrides have attracted a great deal of attention for the use as ultra-hard materials". However the nitrogen-rich compounds have a unique chemistry and may lead to compounds with properties very different from what we might expect[3]. Nobody wants that in attempting to synthesize a superhard material, the chemical reactions would result in an explosive material. Be cautious in testing positive enthalpy compounds for hardness! The mechanical stimulus can trigger a violent result.

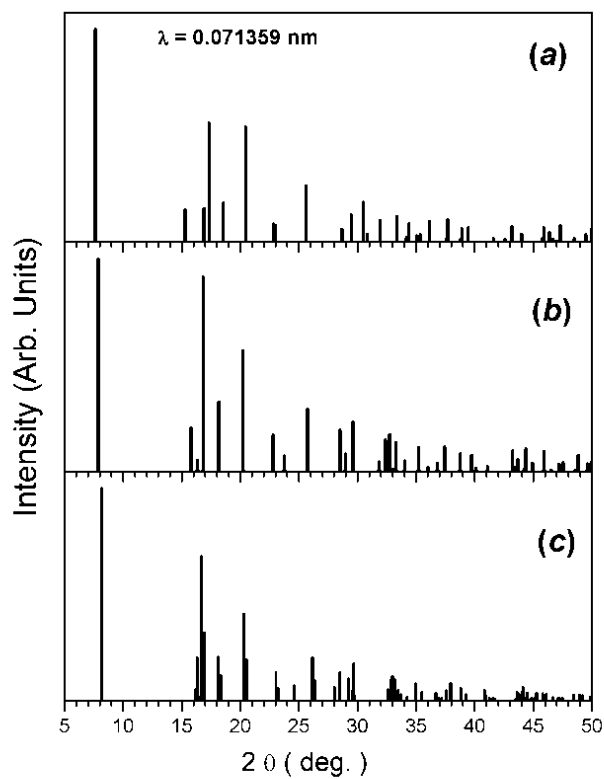

Figure 2.- Generated XRD pattern for $ReN_2$ as it was reported by Kawamura *et. al.*, and patterns as of $ReN_3$ calculated by DFT in the *P63/mmc* space group and in the *Ama2* space group.

Table I. Comparison of Kawamura's experiments with the results given by DFT calculation

| Compound | Figure | Structural Parameters | Equation of state fit parameters[1] |
|---|---|---|---|
| ReN$_2$ | 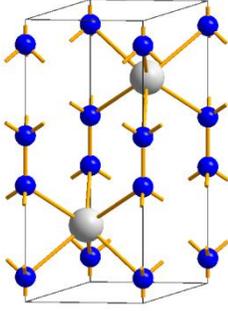 | By Kawamura et. al. (exp.) Space Group: $P6_3/mmc$ (194)<br><br>$a_0 = b_0 = 2.806$,<br>$c_0 = 10.747$<br>$\alpha = \beta = \gamma = 120$<br>Atomic Positions:<br>Re Wyckoff 2$d$<br>(0.66667, 0.33333, 0.25)<br>N Wyckoff 4$e$<br>(0, 0, 0.616) | $V_0 = 75.28$ Å$^3$<br><br>$B_0 = 173$ GPa<br><br>$E_t = -67309.793821$ Ry (calculated by DFT)<br><br>$E_A = +0.55588$ eV atom$^{-1}$<br><br>$E_B = -5.86597$ eV atom$^{-1}$ |
| | | Relaxed by DFT: Space Group: $P6_3/mmc$ (194)<br><br>$a_0 = b_0 = 2.934$,<br>$c_0 = 7.742$<br>$\alpha = \beta = \gamma = 120$<br>Atomic Positions:<br>Re Wyckoff 2$d$<br>(0.66667, 0.33333, 0.25)<br>N Wyckoff 4$e$<br>(0, 0, 0.088355) | $V_0 = 57.70$ Å$^3$<br><br>$B_0 = 350$ GPa<br><br>$dB_0 = 7.0710$<br><br>$E_t = -67310.00987$ Ry<br><br>$E_A = +0.06596$ eV atom$^{-1}$<br><br>$E_B = -6.35589$ eV atom$^{-1}$ |
| ReN$_3$ | 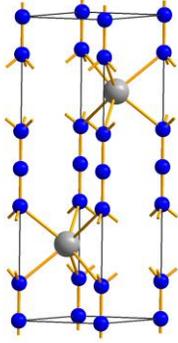 | Space Group: $P6_3/mmc$ (194)<br><br>$a_0 = b_0 = 2.8995$,<br>$c_0 = 10.41$<br>$\alpha = \beta = \gamma = 120$<br><br>Atomic Positions:<br>Re Wyckoff 2$d$<br>(0.66667, 0.33333, 0.25)<br>N Wyckoff 4$e$<br>(0, 0, 0.12365)<br>N Wyckoff 2$a$<br>(0, 0, 0) | $V_0 = 75.79$ Å$^3$<br><br>$B_0 = 293.6299$ GPa<br><br>$dB_0 = 7.09$<br><br>$E_t = -67528.706169$ Ry<br><br>$E_A = +0.64731$ eV atom$^{-1}$<br><br>$E_B = -6.57727$ eV atom$^{-1}$ |
| ReN$_3$ | 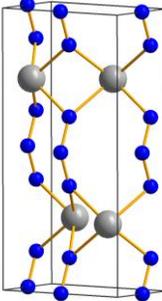 | Space Group: $Am2a$ (40)<br><br>$a_0 = 10.0622$, $b_0 = 5.00002$,<br>$c_0 = 2.95$<br>$\alpha = \beta = \gamma = 90$<br><br>Atomic Positions:<br>Re Wyckoff 4$b$<br>(0.25, 0.66751, 0.30367)<br>N Wyckoff 8$c$<br>(0.12255, 0.00012, 0.31473)<br>N Wyckoff 4$a$<br>(0, 0, 0.48985) | $V_0 = 148.42$ Å$^3$<br><br>$B_0 = 321.9052$ GPa<br><br>$dB_0 = 4.7897$<br><br>$E_t = -67528.791356$ Ry<br><br>$E_A = +0.50243$ eV atom$^{-1}$<br><br>$E_B = -6.72215$ eV atom$^{-1}$ |

[1] $E_t(Re_{HCP} = -31252.13779$ Ry); $E_t(N_2 = -219.04782$ Ry); $E_t(N_{Atomic} = -108.81591$ Ry), values obtained by DFT using the same level of theory than the ReN$_x$ compounds.

Dear Dr. Soto,

Your manuscript, referenced below, has been reviewed for Applied Physics Letters.

"Comment on "Synthesis of rhenium nitride crystal with MoS2 structure" [Appl. Phys. Lett. 100, 251910 (2012)]"
LN12-08824-COMMENT

Attached or included below is the report of the reviewer. In view of his/her recommendations, we cannot accept your paper for publication in this Journal.

Sincerely yours,

…
Associate Editor
APPLIED PHYSICS LETTERS

---
Manuscript #LN12-08824-COMMENT:

Editor's / Reviewer Comments:
Evaluations:
RECOMMENDATION: Reject

Paper Interesting: No
Original Paper: Yes
Sufficient Physics: No
Well Organized: Yes
Clear and Error Free: No
Conclusions Supported: No
Appropriate Title: Yes
Good Abstract: Yes
Satisfactory English: Yes
Adequate References: Yes
Clear Figures: Yes
OVERALL RATING: Poor; not an APL

 (Comments to the Author):

Although the ultimate approval or rejection of the discussion raised by Soto based on DFT calculations requires a more detailed theoretical study, we could not verify the validity of the claims for the following reasons:

1. The compound studied by Kawamura et al. has been synthesized under high pressure-high temperature conditions and thus, is a metastable phase. However, we do not know if Soto has carried out his DFT study under this non-equilibrium condition. We think including these conditions in the calculations may lead to different results that can better match Kawamura et al.'s data.
R.- The volume of Kawamura's proposed structure is larger than the volume of the DFT-relaxed cell. I don't think possible that a structure that was synthesized under pressure can develop a metastable phase that is a "tensioned variation" of the relaxed structure.

2. Since Kawamura et al. imply in their paper that their material is
conductive, which is also supported by their XPS data (asymmetric peaks and
larger than zero intensity at binding energy = 0), we cannot believe that
their material is a compound salt such as the azide suggested by Soto. In
fact, from the micrograph that they have presented in their paper their
material appears to be a layered compound. Hence the as-synthesized material
is likely a layered metallic compound.

R.- As can be see from the calculated density of states of ReN$_3$, this compound is conductive, i.e. the density of states at the Fermi level is not zero, although there is a pseudo-gap. The conductive characteristics of this material should be similar to a semi-metal. Additionally, the ReN$_3$ is similar to ReN$_2$, that is, a layered compound where there are nitrogen layers that are sandwiched between rhenium layers. The ReN$_2$ is with two layers of nitrogen (ReNNRe) while the ReN$_3$ is with three layers (ReNNNRe). In fact, the ionic character is lower in the azide than in the diazenide, and the ionic character of the diazenide is lower than the nitride. The formal charges are:

Azide < diazenide < nitride

$(N_3)^- < (N_2)^{2-} < N^{3-}$

3. There may have been a misunderstanding between an azide and a trinitride;
to our knowledge based on the existing literature, only P-block elements and
1st row transition metals are capable of forming trinitrides. So, we think
forming a trinitride by rhenium would be very unusual. It would be great if
Soto could provide some examples of trinitrides of 2nd and 3rd row transition
metals.

R. The use of the word "trinitride" is incorrect here. As M. Wessel and R. Dronskowski discuss (JACS 132, 2421(2010)) a nitride phase contains by definition isolated nitrogen anions with a formal charge of -3. A trinitride would yield an implausible oxidation state of +9 to rhenium atoms. However this compound contains azide anions $(N_3)^-$ with a formal charge of -1. Azide is the correct nomenclature for ReN$_3$, while diazenide (or pernitride) is the correct nomenclature for ReN$_2$. There are many examples of azide forming metals:

AgN$_3$, Hg$_2$N$_6$, Pd(N$_3$) (W. Beck, et al., Eur. J. Inorg. Chem. 1999, 523-526), barium azide, thallium azide, copper azide, nickel azide, etc. The heavy metal azides are difficult to characterize because they tend to explode. Typically they are very sensitive explosives.

4. We disagree with Soto's last statement about the application of nitrides
as superhard materials. There are several nitrides used in applications such
as cutting tools without any "explosion". A good example is TiN that has been
widely used in tools without any trouble. So, we think Soto should be more
careful with his statements.

R. I'm not talking about the nitrides in the last statement. I was talking about the nitrogen-rich, azide and diazenide, compounds. As clearly say *"Be cautious in testing positive enthalpy compounds for hardness!"*. The titanium nitride is not a positive-enthalpy compound; however the ReN$_2$ and the ReN$_3$ are.

5. There are also some confusing statements in this comment that make little
sense. Here is an example: "As a first test, DFT shows that the structure as
proposed by Kawamura is not in equilibrium, that is, can not be a metastable
phase."

R. The original version was reduced due to limitations of space. The figure 1 was not in the original version. Maybe this is the origin of confusion. As can be see now, the Kawamura structure is not in equilibrium because its derivate of energy is not zero.

In conclusion, it appears that this comment is off base and should be
rejected/withdrawn.
----------------------------------------------------------------